\documentclass[floatfix,twocolumn,nofootinbib,superscriptaddress,a4paper,preprintnumbers]{revtex4-1}

\usepackage{amsmath}
\usepackage{amssymb}
\usepackage{graphicx}
\usepackage[T1]{fontenc}
\usepackage{slashed}
\usepackage[utf8]{inputenc}
\usepackage{xspace}
\usepackage{color}
\usepackage{ulem}
\usepackage{array}

\newcommand{\calO}{{\cal O}}

\begin{document}

%\preprint{KA-TP-26-2017}

\title{Realization of Sneutrino Self-interacting Dark Matter in the Focus Point Supersymmetry}

\author{Bin Zhu}
\email{zhubin@mail.nankai.edu.cn}
\affiliation{Department of Physics, Yantai University, Yantai 264005, P. R. China}

\author{Ran Ding}
\email{dingran@mail.nankai.edu.cn}
\affiliation{Center for High-Energy
Physics, Peking University, Beijing, 100871, P. R. China}

\author{Ying Li}
\email{liying@ytu.edu.cn}
\affiliation{Department of Physics, Yantai University, Yantai 264005, P. R. China}

\begin{abstract}
The Lightest Supersymmetric Particle (LSP) is generally regarded as higgsino in focus point supersymmetry (SUSY). Under such circumstance it leads to bileptino LSP when the MSSM is extended by $U(1)_{B-L}$ gauge group within the framework of Double Focus Point (DFP) supersymmetry. The bileptino is a copy of higgsino whose partner, bilepton, is used to break $U(1)_{B-L}$ gauge symmetry spontaneously. Such scenario, however, is not favored by direct detection since it leads to unacceptable spin-independent cross section when one requires correct self-scattering cross section. We point out that is not necessarily the case even in the presence of light $Z^{\prime}$. The right-handed sneutrino in this model acts as LSP in most of parameter space for non-vanishing soft trilinear coupling $T_{\eta}$. It is thus consistent of the requirement of DFP SUSY without involving any direct detection issue. The corrected relic abundance could be achieved via sneutrino annihilate into pair of bilepton which also serve as light mediator in Self-Interacting Dark Matter (SIDM). Moreover, stringent constraint comes from CMB anisotropies can be evaded by considering retarded decay of right-handed neutrinos. We further stress the need for a large soft trilinear term $T_{\eta}$ in order to generate desirable self-scattering cross section $\sigma/m_{\tilde\nu_R}$ with moderate Yukawa coupling $Y_{\eta}$. The numerical calculation illustrates that SIDM is reliable in our model from the scales of dwarf galaxy to galaxy cluster.
\end{abstract}

\maketitle

% =============================================================================
%\section{Introduction}
%\label{sec:intro}
% =============================================================================
\section{Introduction}
SUSY~\cite{Martin:1997ns} is a fermion-boson symmetry and is an indispensable layout of Coleman-Mandula no-go theorem~\cite{Coleman:1967ad}. Its particle content is larger than the Standard Model (SM) in the sense that additional copied degrees of freedom are required, i.e., superpartners of SM particles. With the existence of superpartners at hand especially the scalar top partner, the quadratic divergence in loop correction of higgs mass is cancelled even for SUSY is soft breaking ~\cite{Nilles:1983ge} and electroweak hierarchy problem is solved naturally. Our expectation of what to find beyond the SM has been deeply shaped by this naturalness arguments, and arguably low energy SUSY
emerged as the primary candidate for BSM physics. In addition,  a discrete parity called $R$-parity is imposed to forbid dangerous proton decay under which each particle bears a quantum number,
\begin{align}
P_R=(-1)^{3(B-L)+2s}\;,
\end{align}
with $B$, $L$ and $s$ are respectively the baryon number, lepton number and spin of particle. This parity automatically guarantees LSP is stable, and could be naturally qualified for a Cold Dark Matter (CDM) candidate~\cite{Bertone:2004pz}. Therefore, the simplest realization of SUSY at electroweak scale, Minimal Supersymmetric SM (MSSM), leading the study of BSM physics during last decades.

However, the situation is challenged by recent experiments. From the perspective of naturalness, the discovery of $125~\text{GeV}$ Higgs boson~\cite{Aad:2012tfa,Chatrchyan:2012xdj} together with non-observation of sparticles in LHC put stringent limits on parameter space of MSSM. This motivates us to consider other possible realization of SUSY. There are in general two different approaches to improve the current issue: adding new fields or introducing  new SUSY breaking scenario. For the first approach, a well-known example is NMSSM~\cite{Ellwanger:1996gw,Ellwanger:2006rm,Ellwanger:2009dp}, which improving fine-tuning through additional tree-level F-term contribution on higgs mass.
On the other hand, the focus point mechanism~\cite{Feng:1999zg,Feng:2013pwa,Feng:2011aa,Feng:2012jfa,Feng:2000bp,
Horton:2009ed,Brummer:2012zc,Yanagida:2013ah,Agashe:1999ct,Draper:2013cka,Ding:2013pya}, who represents second approach, keeping heavy enough sparticles to lift higgs mass while remaining fine-tuning under control. Above two schemes can be even combined in gauge extension SUSY framework, such as $U(1)_{B-L}$ extended MSSM model (BLSSM)~\cite{Un:2016hji,DelleRose:2017ukx,DelleRose:2017smp,Zhu:2017moa,Zhao:2016jcx}  which is the main focus of this paper.

On the dark matter side, neutralino is severely limited by direct, indirect and collider DM searches. Particularly, vast parameter space of neutralino DM has been ruled out by direct detection experiments, even for the popular well-tempered neutralino~\cite{ArkaniHamed:2006mb}. In spite of these constraints, the CDM paradigm itself also encounter difficulty on interpreting the small scale structure observations~\cite{Flores:1994gz,BoylanKolchin:2011de,BoylanKolchin:2011dk}, which is known as the core-vs-cusp problem~\cite{Oh:2010ea} and the too-big-to-fail problem~\cite{Walker:2012td}. SIDM~\cite{Peter:2012jh,Tulin:2012wi,Kaplinghat:2013yxa,Tulin:2013teo,Ko:2014nha,Wang:2014kja,Kang:2016xrm,Tulin:2017ara}  provides an valid solution to reconcile above tension. Due to above intriguing properties, SIDM has been extensively explored and the current status for this issue can be found, for instance, in Ref.~\cite{Tulin:2017ara}. In this scenario, the typical DM self-scattering cross section required for solving small scale structure discrepancy is $\sigma/m_{\rm DM}\sim 1{\rm cm}^2/{\rm g}$ in galaxies, which is much larger than the weak-scale cross section preferred by WIMP CDM model. That strongly invoks a light mediator to enhance cross section via non-perturbative resummation.

Clearly, neutralino fails to standing SIDM. To our knowledge, there is only one  supersymmetric realization for SIDM in the literature, i.e., in the general next-to-minimal supersymmetric SM (GNMSSM)~\cite{Ross:2012nr} model. The underlying reason for choosing GNMSSM rather than scale invariant NMSSM is due to the fact that the vanishing of coupling $\lambda$ in $\lambda S H_u H_d$ interaction (where $S$ denotes singlet chiral superfield) is plausible which makes the singlet sector decouple from visible MSSM sector. Thus the $\{s,\tilde s\}$ sector provides desirable singlino SIDM with ultra-light singlet higgs being force mediator. Inspired by such strategy, we manage to obtain a similar sector of SIDM in the framework of BLSSM: the $U(1)_{B-L}$ extension of MSSM.

The phenomenology of BLSSM with heavy $Z^{\prime}$ (the gauge boson associated with $U(1)_{B-L}$ symmetry) has been extensively explored~\cite{O'Leary:2011yq,Hirsch:2011hg,Basso:2012gz,Un:2016hji,DelleRose:2017ukx,
DelleRose:2017smp}, and case for light $Z^{\prime}$~\cite{Komachenko:1989qn} is also interesting since it can account for novel ${\rm Be}$ anomaly~\cite{Feng:2016jff,Feng:2016ysn,Gu:2016ege,Seto:2016pks}. Notice that light $Z^{\prime}$ causes similar little hierarchy problem as in MSSM which motivating  the proposal of double focus point (DFP) mechanism~\cite{Zhu:2017moa} to solve this problem. Within the framework of BLSSM with DFP, a viable choice for SIDM candidate seems to be bileptino with light $Z^{\prime}$ in $\{Z^{\prime}, \tilde\eta\}$ sector as a mediator. In this paper, however, we prove this recipe will lead to a contradiction between requirements of SIDM and limits from direct detection experiments for bileptino LSP. This is mainly because $Z^{\prime}$ could couple to quarks in terms of non-vanishing gauge interaction, thus the $\{Z^{\prime}, \tilde\eta\}$ sector does not hide itself from MSSM again. Alternatively, we propose right-handed sneutrino as SIDM candidate. The interesting point is that, in DFP, sneutrino LSP requires large trilinear soft term $T_{\eta}$ which is also the necessary condition of SIDM. We therefore obtain a natural SIDM with $\{\tilde\nu_1, \eta\}$ sector where right-handed sneutino LSP $\tilde\nu_1$ and bileptons $\eta$ are serve as SIDM and force mediator, respectively.

The rest of paper is organized as follows: in section~\ref{sec:model} we layout  BLSSM contents which are necessary for SIDM calculation. In particular, we illustrate quantitatively why that bileptino SIDM is not viable. In section~\ref{sec:LSP}, we show the analytical derivation of sneutrino LSP from DFP consideration. Furthermore, the parameter space of sneutrino LSP is discussed in detail. In section~\ref{sec:SIDM}, we perform a numerical calculation of SIDM beyond Born limit where the solution of schrodinger equation at $r\rightarrow\infty$ captures the properties of elastic self-scattering. The relevant DM properties such as relic abundance involving Sommerfeld Enhancement (SE) effect and direct detection are also considered. We finally conclude in section~\ref{sec:conclusion}.

\section{The BLSSM Description}
\label{sec:model}
In BLSSM, the chiral superfields are extended by a pair of bileptons  ($\hat \eta_{1}, \hat \eta_2$) and three generations of right-handed neutrino  $\hat\nu_{R_i}$. The neutrino mass generation and hierarchy can be obtained via inverse-seesaw mechanism~\cite{Bazzocchi:2009kc} or canonical type-I seesaw induced by SUSY breaking~\cite{ArkaniHamed:2000bq}. We ignore neutrino mass issue in this paper and concentrate on SIDM in DPF SUSY. The complete particle contents and charge assignments are listed in table~\ref{tab:gauge} and \ref{tab:chiral}.
\begin{table}
\begin{center}
\begin{tabular}{|c|c|c|c|c|c|}
\hline \hline
Superfield & Spin \(\frac{1}{2}\) & Spin 1 & Gauge group & Coupling\\
\hline
$\hat{B}$ & \(\lambda_{\tilde{B}}\) & $B$ & $U(1)_Y$ & $g_1$ \\
$\hat{W}$ & \(\lambda_{\tilde{W}}\) & $W$ & $SU(2)_L$ & $g_2$ \\
$\hat{g}$ & \(\lambda_{\tilde{g}}\) & $g$ & $SU(3)_c$ & $g_3$ \\
$\hat{B}^\prime$ & \(\lambda_{\tilde{B}^\prime}\) & $B^\prime$ & $U(1)_{B-L}$ & $g_B$ \\
\hline \hline
\end{tabular}
\end{center}
\caption{
Vector superfields of the BLSSM and corresponding gauge couplings.
\label{tab:gauge}}
\end{table}
\begin{table}
\begin{center}
\begin{tabular}{|c|c|c|c|c|c|}
\hline \hline
Superfield & $N_G$ & $U(1)_Y\otimes\, SU(2)_L\otimes\, SU(3)_c\otimes\, U(1)_{B-L}$ \\
\hline
$\hat Q$  & 3 & $\frac{1}{6}\otimes\,{\bf 2}\otimes\,{\bf 3}\otimes\,\frac{1}{6}$ \\
$\hat U$ & 3 & $-\frac{2}{3}\otimes\,{\bf 1}\otimes\,{\bf \overline{3}}\otimes\,-\frac{1}{6}$ \\
$\hat D$ & 3 & $\frac{1}{3}\otimes\,{\bf 1}\otimes\,{\bf \overline{3}}\otimes\,-\frac{1}{6}$ \\
$\hat L$ & 3 & $-\frac{1}{2}\otimes\,{\bf 2}\otimes\,{\bf 1}\otimes\,-\frac{1}{2}$ \\
$\hat E$ & 3 & $1\otimes\,{\bf 1}\otimes\,{\bf 1}\otimes\,\frac{1}{2}$ \\
$\hat \nu_R$  & 3 & $0\otimes\,{\bf 1}\otimes\,{\bf 1}\otimes\,\frac{1}{2}$ \\
$\hat H_u$  & 1 & $\frac{1}{2}\otimes\,{\bf 2}\otimes\,{\bf 1}\otimes\,0$\\
$\hat H_d$  & 1 & $-\frac{1}{2}\otimes\,{\bf 2}\otimes\,{\bf 1}\otimes\,0$ \\
$\hat \eta_1$ & 1 & $0\otimes\,{\bf 1}\otimes\,{\bf 1}\otimes\,-1$ \\
$\hat \eta_2$ & 1 & $0\otimes\,{\bf 1}\otimes\,{\bf 1}\otimes\,1$ \\
\hline \hline
\end{tabular}
\end{center}
\caption{
Chiral superfields of the
BLSSM and their charges under $U(1)_Y\otimes\, SU(2)_L\otimes\, SU(3)_c\otimes\, U(1)_{B-L}$ gauge group.\label{tab:chiral}}
\end{table}
The relevant superpotential of BLSSM is given as
\begin{align}
W = & Y^{ij}_u \hat U_i \hat Q_j \hat H_u\,- Y^{ij}_d \hat D_i \hat Q_j \hat H_d\,- Y^{ij}_e \hat E_i \hat L_j \hat H_d\, +\mu \hat H_u \hat H_d \nonumber\\
   & Y^{ij}_{\eta} \hat \nu_{Ri} \hat \eta_1 \hat \nu_{Rj}\,+Y^{ij}_\nu \hat L_i \hat H_u \hat \nu_{Rj}\, - {\mu_\eta} \hat \eta_1 \hat \eta_2\;,
\label{eq:st}
\end{align}
where $i,j$ denote family indices and all color and isospin indices are suppressed. The first line in equation~(\ref{eq:st}) represents the conventional Yukawa interaction as well as $\mu$ term in MSSM; while the second line stands for the additional interactions induced by extended gauge group $U(1)_{B-L}$. The corresponding soft-breaking terms are given as
\begin{align}
\mathcal{L}_{\text{BLSSM}} &= \mathcal{L}_{\text{MSSM}} - {M}_{B B^\prime} \lambda_{\tilde{B}} \lambda_{\tilde{B}^\prime} - \frac{1}{2} {M}_{B^\prime} \lambda_{\tilde{B}^\prime} \lambda_{\tilde{B}^\prime} \nonumber \\
& - m_1^2 |\eta_1|^2 - m_2^2 |\eta_2|^2  - {m_{\nu,ij}^{2}} (\tilde{\nu}_{Ri}^c)^* \tilde{\nu}_{Rj}^c   \nonumber \\
&- B_{\mu_\eta}\eta_1 \eta_2 + T^{ij}_{\nu}  H_u \tilde{\nu}_{Ri}^c \tilde{L}_j + T^{ij}_\eta \eta_1 \tilde{\nu}_{Ri}^c \tilde{\nu}_{Rj}^c\;.
\end{align}
We will see that $Y_{\eta}$ and $T_{\eta}=Y_{\eta} A_{\eta}$ play a crucial role in determining SIDM cross section. Notice that $T_{\eta}$ leads to intrinsic mixing in right-handed sneutrino sector, while $T_{\nu}$ leads to left-right mixing between left-handed sneutrino sector and right-handed one. The conventional gravity mediation indicates that $T_{\nu}=Y_{\nu}A_{\nu}$, thus the left-right mixing can be neglected due to the vanishing $Y_{\nu}$. There is only right-right mixing left in this BLSSM, and mass splitting between the CP even and odd part yields
\begin{align}
\tilde\nu_L=\frac{1}{\sqrt{2}}\left(i \sigma_{L} + \phi_{L} \right)\;, \tilde\nu_R=\frac{1}{\sqrt{2}}\left(i \sigma_{R} + \phi_{R} \right)\;,
\end{align}
where $\{\phi_L, \phi_R\}$ are mixed into CP-even sneutrinos, and $\{\sigma_L,\sigma_R\}$ are mixed into CP-odd sneutrinos. Through Higgs states and bileptons receiving vacuum expectation values (VEVs), the electroweak and $U(1)_{B-L}$ symmetry are spontaneously broken into $U(1)_{e.m.}$. After symmetry breaking, the complex scalars are parameterized as
\begin{eqnarray}
H_d^0 &=& \frac{1}{\sqrt{2}} \left(i \sigma_{d} + v_d  +  \phi_{d} \right)\;,
H_u^0 = \frac{1}{\sqrt{2}} \left(i \sigma_{u} + v_u  +  \phi_{u} \right)\;,\nonumber\\
\eta_1 &=& \frac{1}{\sqrt{2}} \left(i \sigma_1 + v_1 +  \phi_1 \right)~, \, \eta_2 = \frac{1}{\sqrt{2}} \left(i \sigma_2 + v_2  +  \phi_2 \right)\;.
\end{eqnarray}
in analogous with $\tan\beta$ in MSSM, we here denote the ratio of the two bilepton VEVs as $\tan\beta^\prime=v_1/v_2$. The CP-even higgs sector is composed from mixing of gauge eigenstates $\{\phi_1,\phi_2,\phi_u,\phi_d\}$. For CP-odd higgs, two of the four gauge eigenstates survive after gauge symmetry,
\begin{align}
m_{A^0}^2=\frac{2B_{\mu}}{\sin2\beta}\;, m_{A_{\eta}^0}^2=\frac{2B_{\mu_{\eta}}}{\sin2\beta^{\prime}}\;.
\end{align}
Since $B_{\mu}$ and $B_{\mu_{\eta}}$ are not related to DFP, they are relatively heavy so as not to be a candidate of light force carrier. Meanwhile CP-even Higgs mass especially $\phi_1$ is quantified as $m_{\eta_1}^2$ which is very tiny in DFP, thus it naturally leads to light force carrier. For simplicity, we define $\eta_1$ as a lightest CP-even Higgs in order to recall that it comes mostly from the gauge eigenstate $\eta_1$.

A special property of BLSSM is that it gives rise to gauge-kinetic mixing term with two Abelian gauge groups $U(1)_{B-L}$ and $U(1)_Y$ via
\begin{align}
\mathcal{L}=\frac{1}{4}\xi F_{\mu\nu}^{\rm{B-L}}F^{\mu\nu}\;,
\end{align}
where $\xi$ is function of $\tilde{g}$ in equation~(\ref{eq:mix}). Even absent at tree level, it will be reintroduced via loop correction with running effect of Renormalization Group Equation (RGE). In terms of a triangle form of the gauge coupling matrix~\cite{Fonseca:2011vn}, the bilepton contributions to the $Z$ mass vanish:
\begin{equation}
\left(\begin{array}{cc} g_{YY} & g_{YB} \\ g_{BY} & g_{BB}   \end{array} \right) \to  \left(\begin{array}{cc} g_1 & \tilde{g} \\ 0 & g_B \end{array} \right)\;,
\label{eq:mix}
\end{equation}
and the gauge couplings are related by~\cite{Chankowski:2006jk}:
\begin{eqnarray}
\label{eq:Triangle1}
g_1 &=& \frac{g_{YY} g_{BB} - g_{YB} g_{BY}}{\sqrt{g_{BB}^2 + g_{BY}^2}}\;, \nonumber\\
\tilde{g} &=& \frac{g_{YB} g_{BB} + g_{BY} g_{YY}}{\sqrt{g_{BB}^2 + g_{BY}^2}};, \nonumber\\
g_B &=& \sqrt{g_{BB}^2 + g_{BY}^2}\;.
\label{eq:triangle}
\end{eqnarray}
In addition, after electroweak and $U(1)_{B-L}$ breaking, the gauge-kinetic mixing further induces a mixing between the neutral SUSY particles from the MSSM and from the new sector, resulting seven neutralinos in this model,
\begin{align}
\chi_1^0 &= c_{\tilde B}\tilde B +c_{\tilde W} \tilde W +c_{\tilde H_u}\tilde H_u+
c_{\tilde H_d}\tilde H_d\nonumber\\
              &+c_{\tilde\eta_1}\tilde\eta_1 +c_{\tilde\eta_2}\tilde\eta_2
                +c_{\tilde \tilde B^{\prime}}\tilde B^{\prime}\;.
\end{align}
The LSP is then determined by the mixing coefficients $c_i$. For instance, in conventional MSSM with focus point SUSY, higgsino is LSP with $c_{\tilde\eta_1}\sim c_{\tilde\eta_2}\sim 0.5$. DM in BLSSM with heavy $Z^{\prime}$ has been discussed in~\cite{Basso:2012gz}. While BLSSM with light $Z^{\prime}$ is different and introduces similar little hierarchy problem with MSSM from the perspective of tadpole equation:
\begin{align}
m_Z^2\sim -2m_{H_u}^2-2\mu^2~,\; m_{Z^{\prime }}^2\sim -m_\eta^2-\mu_\eta^2\;.
\label{eqn:zp}
\end{align}
From above equation, one can see that for heavy $Z^\prime$ around TeV, there is no need to worry about the fine-tuning issue. However, a very light $Z^\prime$ is still allowed for tiny $g_{BL}$ \cite{Komachenko:1989qn}, which forces us to consider the fine-tuning seriously. In order to retain naturalness, $m_{H_u}^2$ and $m_{\eta}^2$ should vanish at electroweak scale simultaneously through RGE running effect. Such a mechanism is called DFP SUSY,
\begin{align}
\left[\begin{array}{c}
                m_{H_u}^2[Q_{\text{GUT}}] \\
                m_{q}^2[Q_{\text{GUT}}] \\
                m_{u}^2[Q_{\text{GUT}}] \\
                A_{t}^2[Q_{\text{GUT}}]
              \end{array}\right]
            &  =\left[\begin{array}{c}
                m_{0}^2 \\
                m_{0}^2+\kappa _0^{\prime}-\frac{2 \kappa
   _{12}}{3} \\
                m_{0}^2-\kappa _0^{\prime}-\frac{4 \kappa
   _{12}}{3} \\
                6\kappa_{12}
              \end{array}\right]\nonumber\\
&\rightarrow\nonumber\\
\left[\begin{array}{c}
                m_{H_u}^2[Q_{\text{SUSY}}] \\
                m_{q}^2[Q_{\text{SUSY}}] \\
                m_{u}^2[Q_{\text{SUSY}}] \\
                A_{t}^2[Q_{\text{SUSY}}]
              \end{array}\right]
              &=\left[\begin{array}{c}
                0 \\
                \frac{m_{0}^2}{3}+\kappa _0^{\prime}-\frac{2 \kappa
   _{12}}{5} \\
                \frac{2m_{0}^2}{3}+\kappa _0^{\prime}-\frac{4 \kappa
   _{12}}{5} \\
                \frac{2}{3}\kappa_{12}
              \end{array}\right]\;,
\label{eqn:focus1}
              \end{align}
and
\begin{align}
\left[\begin{array}{c}
                m_{\eta_1}^2[Q_{\text{GUT}}] \\
                m_{\nu_R}^2[Q_{\text{GUT}}] \\
                A_{\eta}^2[Q_{\text{GUT}}
              \end{array}\right]
 &  =\left[\begin{array}{c}
                m_{0}^2 \\
                \frac{43}{54}m_0^2-\frac{7}{20}\epsilon_{28} \\
                7\epsilon_{28}
              \end{array}\right]\nonumber\\
 &\rightarrow\nonumber\\
    \left[\begin{array}{c}
                m_{\eta_1}^2[Q_{\text{SUSY}}] \\
                m_{\nu_R}^2[Q_{\text{SUSY}}] \\
                A_{\eta}^2[Q_{\text{SUSY}}]
              \end{array}\right]
             & =\left[\begin{array}{c}
                0 \\
               \frac{7}{54}m_0^2-\frac{7}{20}\epsilon_{28} \\
                \frac{7}{100}\epsilon_{28}
              \end{array}\right]\;.
\label{eqn:focus3}
          \end{align}

At first glance, since the $m_{\eta}^2$ becomes tiny through DFP mechanism, the corresponding $\mu_{\eta}$ is small too. The bilpetino $\{\tilde\eta_{1,2}\}$ thus become LSP when $Z^{\prime}$ is much lighter than $Z$ boson. We also mention that once the gauge kinetic term $\tilde g$ is tiny compared with $g_{BL}$, the $U(1)_{B-L}$ sector then decoupled from MSSM sector which is the main point of obtaining SIDM in BLSSM. Furthermore, the hidden sector requirement in~\cite{Zhu:2017moa} is also the necessary condition of DFP SUSY.

It appears that the bileptino LSP with ultra-light $Z^{\prime}$ being mediator i.e. $\{Z^{\prime},\tilde\eta_1,\tilde\eta_2\}$ sector can furnish itself as SIDM without any fine-tuning issue. Nevertheless, it is not true when we consider the direct detection. As we know, LSP is majorana type neutralino in MSSM which undergos spin-dependent scattering off nuclei via the exchange of $Z$ boson. In the case of bileptino in BLSSM, the interaction mediated by light $Z^{\prime}$ is however vector like type which leads to a spin-independent interaction. Such an interaction can be constrained by the recent direct detection limits such as LUX~\cite{Akerib:2016vxi}, XENON1T~\cite{Aprile:2018dbl} and PandaX-II~\cite{Cui:2017nnn} experiments. The spin-independent scattering cross section can be written in terms of the spin-averaged squared matrix element,
\begin{align}
\sigma_{\text{SI}}=\frac{\mu^2}{16\pi m_{\chi}^2 m_A^2}\left(\frac{1}{4}\sum_{s} |\mathcal{M}|^2\right)\;,
\end{align}
where $\mu$ is the reduced mass and $m_A$ is the mass of target nucleus. The effective operator for spin-independent DM scattering through $Z^{\prime}$ exchange can be written as,
\begin{align}
\mathcal{O}_{Z^{\prime}}=\lambda_{\chi Z^{\prime}}\lambda_{q Z^{\prime}}\frac{1}{m_{Z^{\prime}}}(\bar\chi\gamma^{\mu}\chi)(\bar q\gamma_{\mu}q)\;,
\end{align}
with $\lambda_{q Z^{\prime}}$ is the vertex between $Z^{\prime}$ and quarks which is determined by $U(1)_{B-L}$ gauge coupling $g_{BL}$. The coupling $\lambda_{\chi Z^{\prime}}$ is given explicitly as
\begin{align}
\lambda_{\chi Z^{\prime}}&=g_{BL}(|N_{15}|^2-|N_{16}|^2)\;,
\end{align}
with $N_{15}$ and $N_{16}$ are the same as $c_{\eta_{1,2}}$ and stands for bileptino fraction in LSP. Therefore we find the direct detection constraint can be evaded when bileptino is composed with equivalent fraction of $\tilde\eta_1, \tilde\eta_2$, i.e., $N_{15}=N_{16}=0.5$ or very small $g_{\rm{BL}}$. The purity of Dirac bileptino seems to escape from direct detections for almost cancelled $\lambda_{\chi Z^{\prime}}$. However, under these circumstance it leads to vanishing SIDM scattering cross section since $\lambda_{\chi Z^{\prime}}$ also determines the self-interacting scattering process. In turn, large self-interacting cross section results in large spin-independent cross section. This is the biggest challenge of BLSSM for bileptino having the same coupling in direct detection and SIDM.

The conventional approach for escaping direct detection constraint is to break Dirac-type property of bileptino. The degeneracy between bileptinos is broken by the non-vanishing expectation value of $\eta_1,\eta_2$. If the mass spitting is above keV which is the threshold of direct detection, i.e., the sensitivity of dark matter nucleus collision, there is only one Majorana bileptino left in the direct detection, the spin-independent scattering mediated by $Z^{\prime}$ boson turns out to be spin-dependent type whose constraint is relatively relaxed. Nevertheless, if the mass difference $\delta$ is as small as $\mathcal{O}(100)\,\text{keV}$, the inelastic scattering process occurs and reintroduces the constraint from direct detection~\cite{TuckerSmith:2001hy,TuckerSmith:2004jv,Chang:2008gd}. In BLSSM, the mass splitting~\footnote{The derivation of mass splitting could be found in appendix~\ref{app:splitting}. The inelastic spin-independent constraints would be studied in up-coming paper.} between bileptinos is approximately proportional to $m_{Z^{\prime}}^2/2M_{B^{\prime}}$, which is below several hundred keV thus constrained by inelastic scattering again. Finally, we conclude that bileptino $\tilde\eta_{1,2}$ can not be regarded as SIDM in BLSSM unless very fine-tuned spectrum is imposed.

\section{Sneutrino LSP Consideration}
\label{sec:LSP}

The inconsistency between SIDM and BLSSM is originated from the wrong illustration that DFP SUSY yields a higgsino-like LSP, i.e., bileptino in BLSSM. We point out that it is not necessarily the case even in the presence of light $Z^{\prime}$.  From equation~(\ref{eqn:focus3}), it is easy to find that $m_{\nu_R}^2$ can be as small as $m_{\eta_1}$ when large $A_{\eta}$ is given at the boundary. The sneutrino, especially right-hand one could also provide a plausible candidate of SIDM.  In particular the sneutrino mass can even vanish at one loop when $\epsilon_{28}\sim 10/27 m_0^2 $.

After neglecting the kinetic mixing effect $\tilde g$ and setting $Y_{\nu}$ to be zero, we obtain the mass of right-handed sneutrinos in a compact form~\cite{OLeary:2011vlq}:
\begin{align}
m_{\tilde\nu^+}^2=&m_{\nu_R}^2+m_{Z^{\prime}}^2\left(\frac{1}{4}\cos2\beta^{\prime}+\frac{2Y_{\eta}^2}{g_{BL}^2}\right)\nonumber\\
&+m_{Z^{\prime}}\frac{\sqrt{2}Y_{\eta}}{g_{BL}}\left(A_{\eta}\sin\beta^{\prime}-\mu_{\eta}\cos\beta^{\prime}\right)\;,\nonumber\\
m_{\tilde\nu^{-}}^2=&m_{\nu_R}^2+m_{Z^{\prime}}^2\left(\frac{1}{4}\cos2\beta^{\prime}+\frac{2Y_{\eta}^2}{g_{BL}^2}\right)\nonumber\\
&-m_{Z^{\prime}}\frac{\sqrt{2}Y_{\eta}}{g_{BL}}\left(A_{\eta}\sin\beta^{\prime}-\mu_{\eta}\cos\beta^{\prime}\right)\;.
\label{eqn:mass}
\end{align}
Here the sign $+$ ($-$) stands for CP-even (CP-odd) sneutrino. In DFP SUSY, the $m_0$ and $A_{\eta}$ are related as $A_{\eta}\sim\pm 0.18 m_0$ in order to guarantee $m_{\nu_R}$ to be LSP. The typical value of $m_0$ in DFP is $(5-8)\times 10^3~\text{GeV}$ in order to lift Higgs mass to $125$ GeV. Large $m_0$ however does not result in fine-tuning in the context of DFP SUSY. Thus, $A_{\eta}$ could be $\pm (900-1400)~\text{GeV}$ and leads to totally different result depending on the sign of $A_{\eta}$. The combination of $A_{\eta}\sin\beta^{\prime}-\mu\cos\beta^{\prime}$ further determines whether or not CP-even sneutrino is LSP. If $A_{\eta} \tan\beta^{\prime}\sim \mu_{\eta}$, we obtain a nearly degenerated right-hand sneutrino sector where the calculation of SIDM and SE must consider the inelastic scattering with excited state~\cite{Slatyer:2009vg,Das:2016ced,Zhang:2016dck}. However large $\tan\beta^{\prime}\sim 10$ is necessary condition for DFP, which leads to large $\mu_{\eta}$ in degenerated sneutrino sector. That is unnatural so that we ignore it in this paper. That is to say the $\mu_{\eta}$ is suppressed by not only itself but small $\cot\beta^{\prime}$. The most important and relevant term for LSP determination is the sign of $A_{\eta}$.

\begin{figure} [htbp]
\begin{center}
\includegraphics[width=0.4\textwidth]{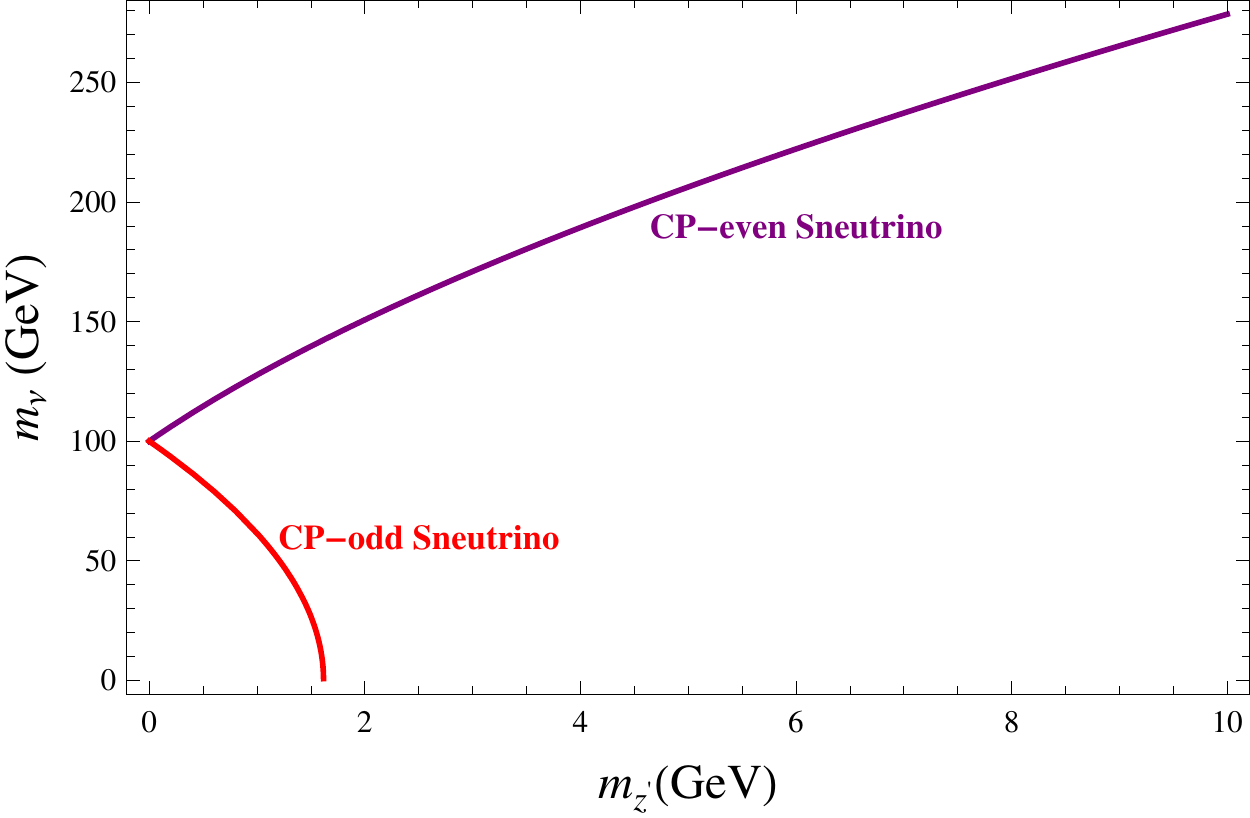}
\includegraphics[width=0.4\textwidth]{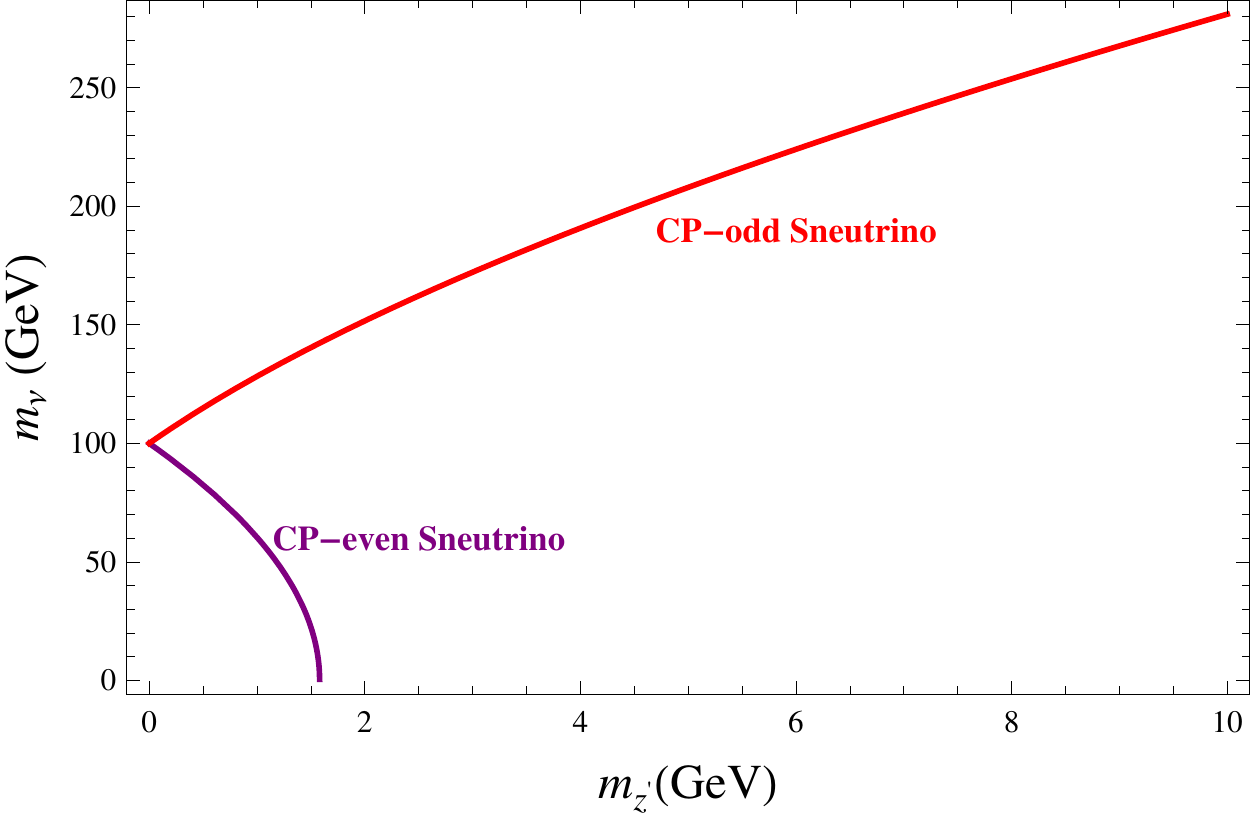}
\end{center}
\caption{The dependence of sneutrino mass on $m_{Z^{\prime}}$ is given explicitly with $m_{\tilde\nu_R}=100~\text{GeV},\,\tan\beta^{\prime}=10,\, Y_{\eta}=0.5,\, A_{\eta}=\pm~900~\text{GeV},\,\mu_{\eta}
=100~\text{GeV},\, g_{BL}=0.1$. The positive $A_{\eta}$ corresponds to CP-odd sneutrino LSP, while negative $A_{\eta}$ corresponds to CP-even sneutrino LSP.}
\label{fig:lightzp}
\end{figure}

In figure~\ref{fig:lightzp}, we plot the behavior of sneutrino as a function of $Z^{\prime}$ mass after setting $m_{\tilde\nu_R}=100~\text{GeV},\,\tan\beta^{\prime}=10,\, Y_{\eta}=0.5,\, A_{\eta}=\pm 900~\text{GeV},\,\mu_{\eta}
=100~\text{GeV},\, g_{BL}=0.1$. The upper panel of  figure~\ref{fig:lightzp} indicates that CP-odd sneutrino is LSP for positive $A_{\eta}$. This is because $A_{\eta}$ drives CP-odd sneutrino lighter than $100~\text{GeV}$ bileptinos. On the opposite, negative $A_{\eta}$ leads to CP-even sneutrino LSP. We should mention that the $Z^{\prime}$ mass must be $2~\text{GeV}$ in order to evade the dangerous tachyon problem of sneutrinos.

The only tree-level interactions responsible for dark matter annihilation are higgs mediator and $Z^{\prime}$ mediator. It is interesting to note that $Z^{\prime}$ must be removed since it couples to one CP-even and one CP-odd sneutrino at a time~\cite{Basso:2012gz} that the coupling itself is always off-diagonal i.e., $Z^{\prime}\tilde\nu_1^{+}\tilde\nu_1^{-}$. The inelastic scattering is equivalent to large $\mu_{\eta}$ in order to cancel $A_{\eta}$ contribution. Therefore it ultimately leads to large fine-tuning. As a result we only consider the higgs mediator which is now light bilepton sector for relatively heavy LSP sneutrino $\tilde\nu_1$. The effective lagrangian of hidden sector is then given as
\begin{align}
\mathcal{L}_{\tilde\nu_1\tilde\nu_1 \eta_1}&=\frac{1}{2}\partial^{\mu}\eta_1\partial_{\mu}\eta_1-\frac{m_{\eta}^2}{2}\eta_1^2+\frac{1}{2}\partial^{\mu}\tilde\nu_1\partial_{\mu}\tilde\nu_1-\frac{m_{\tilde\nu_1}}{2}\tilde\nu_1^2\nonumber\\
&-\frac{\lambda}{2}\eta_1\tilde\nu_1^2 -\frac{Y_{\eta}}{2}\eta_1\nu_R^2\;,
\label{eqn:lag}
\end{align}
where $\lambda$ is the coupling constant between the two $\tilde\nu_1$ and $\eta_1$ which is used to mediate long-range interaction for SIDM and SE. The last term represents the interaction between bilepton mediator and right-handed neutrinos which comes from superpotential~(\ref{eq:st}). All the parameters in equation~(\ref{eqn:lag}) are the function of SUSY breaking effects or superpotential couplings. Depending on whether sneutrino LSP is CP-even or CP-odd, the coupling constant $\lambda$ is read as
\begin{align}
\lambda_{\tilde\nu_1\tilde\nu_1 \eta_1}^{+/-}&=\mp\sqrt{2} T_{\eta}-4v_{\eta}Y_{\eta}^2\nonumber\\
  &=\mp\sqrt{2} A_{\eta}Y_{\eta}-4\frac{m_{z^{\prime}}}{g_{BL}}\sin\beta^{\prime}Y_{\eta}^2\;.
\label{eqn:lam}
\end{align}
The lagrangian in equation~(\ref{eqn:lag}) describes the relevant interactions of relic density and direct/indirect detection for sneutrino DM which have been discussed in~\cite{DelleRose:2017uas}. The crucial difference is that in this paper we realize the SIDM in sneutrino sector by retaining naturalness.

There is no severe direct detection constraint for right-handed sneutrino DM because the the mediator bilepton has no tree level couplings to quarks. However, direct detection constraint will be reintroduced if bilepton mediator rapidly decays into SM final states before BBN era~\cite{Kaplinghat:2013yxa,DelNobile:2015uua,Bernal:2015ova,Kainulainen:2015sva}, and bileptino DM seems to encounter same obstacle. Such a dilemma can be solved easily in the framework of BLSSM where the mediator dominately decay into right-handed neutrinos~ $\eta_1\rightarrow\nu_R\nu_R$ with the decay width,
\begin{align}
\Gamma[\eta_1\rightarrow\nu_R\nu_R]=Y_{\eta}^2\frac{\sqrt{m_{\eta_1}^2/4-m_{\nu_R}^2}}{2\pi}\;.
\label{eqn:BBN}
\end{align}
Equation~(\ref{eqn:BBN}) there sets the lower bound of $Y_{\eta}$. The dominant channel for relic density and indirect detection is bilepton final states via $\tilde\nu_1\tilde \nu_1\rightarrow \eta_1\eta_1$ through t-channel and u-channel annihilation~\cite{Pospelov:2007mp} plus subsequent decay of bileptons.

At first sight, the total process is s-wave dominated and independent of velocity of DM. However it is not true when we consider the long-range Yukawa potential induced by infinite exchange of bilepton mediators. It modifies the wave-function of the annihilating DM pair at origin thus greatly alters the calculation of relic density and indirect detections. When only one partial wave is dominated in annihilation process, the SE effect can be factorized as
\begin{align}
\sigma_{\text{tot}}=S\sigma_{0}=\left|\psi(0)\right|^2\sigma_{0}\;.
\end{align}
Where $S$ is the SE factor. The impact of SE effect on relic abundance is encoded in the Boltzman equation,
\begin{align}
\frac{dY}{dx}=-\sqrt{\frac{\pi}{45}}\frac{g_{\star,\text{eff}}^{1/2} M_{Pl}m_{\tilde\nu_1}
\langle\sigma v_{\text{rel}}\rangle}{x^2}(Y^2-Y_{\text{eq}}^2)\;,
\end{align}
where $Y=n/s$ with $s=(2\pi^2/45)g_{\star s}T^3$ being the entropy density of the universe. Moreover $g_{\star,\text{eff}}$ and $g_{\star,s}$ are the energy and entropy degrees of freedom, respectively. The most important quantity for capturing particle physics input is thermal average cross section,
\begin{align}
\langle\sigma v_{\text{rel}}\rangle &=\int (\sigma v_{\text{rel}})_{\text{tot}} f(v_1)f(v_2)d^3 v_1 d^3 v_2\nonumber\\
& =\int S_0(\sigma v_{\text{rel}})_{0} f(v_1)f(v_2)d^3 v_1 d^3 v_2\;.
\end{align}
The factor $S_0$ encapsulate the effect of long-range interaction which should be obtained by solving the radial part of schrodinger equation at origin. Here we mention that the subscript $0$ in $S_0$ corresponds to the s-wave contribution where only $l=0$ is considered  though there were semi-analytical studies for higher waves~\cite{Cassel:2009wt,Iengo:2009ni,ElHedri:2016onc}. This is mainly because higher partial wave contributions are suppressed by velocity seriously.

Therefore the total cross section is actually velocity-dependent which is consistent with the basic requirement of SIDM in order to be consistent with astrophysical observations. At low velocity the cross section is enhanced at large extent. Thus it put strong constraints over the SIDM parameter space which will be discussed in next section. In particular CMB reionization at ultra-low velocity almost excludes the whole parameter space for s-wave annihilation SIDM~\cite{Bringmann:2016din}. This is the biggest constraint that we must consider in computing SIDM.

\section{Self-Interacting DM Calculation}
\label{sec:SIDM}
The merit of SIDM is the relevant quantity $\sigma/m_{\tilde\nu_1}$ which undergoes velocity depenendent cross sections: for dwarf galaxies,  $\sigma/m_{\tilde\nu_1}\sim 0.1-10~{\rm cm}^2~{\rm g}^{-1}$ for velocity around $\calO(10)~{\rm km}~{\rm s}^{-1}$ which is sufficient to solve core-vs-cusp and too-big-to-fail problems. While the constraints from Milk way sized galaxies and galaxy clusters require smaller $\sigma/m_{\tilde\nu_1}$ with $0.1-1~{\rm cm}^2~{\rm g}^{-1}$~\cite{Kaplinghat:2015aga}.

Since we are interested in the process with energy transfer in SIDM, the conventional total cross section $\sigma_{\text{tot}}=\int d\Omega d\sigma/d\Omega$ is not suitable for estimating the process without energy transfer. Alternatively, we adapt two representative cross sections which have been discussed in plasma literature: one is transfer cross section $\sigma_T$, the other is viscosity cross section $\sigma_V$,
\begin{align}
\sigma_T=\int d\Omega (1-\cos\theta)\frac{d\sigma}{d\Omega}\;,\quad
\sigma_V=\int d\Omega\sin^2\theta\frac{d\sigma}{d\Omega}\;.
\end{align}
The transfer cross section $\sigma_T$ is used to regulate the forward scattering in terms of weight $(1-\cos\theta)$, which is usually used to study Dirac fermion or complex scalar field. Meanwhile viscosity cross section $\sigma_V$ regulates forward and backward scattering simultaneously in terms of the weight $\sin^2\theta$. Thus it is suitable for studying Majorana fermion or real scalar field DM.  In our paper the LSP is real scalar particle $\tilde\nu_1$ and we use $\sigma_V$ throughout the paper.

The desirable cross section for SIDM is far larger than typical WIMP freeze-out cross section which strongly suggests the existence of a dark hidden sector~\cite{Pospelov:2008jd,Boddy:2014yra} i.e. sneutrino LSP with bilepton mediator. We are now considering the interactions between two slowly moving CP-odd sneutrino DM mediated by light force mediator $\eta$. Since the DM particles are non-relativistic,  the dominant process is the one exchanging multiple scalars $\eta_1$ while undergoing either an annihilation for relic abundance or scattering process for self-interaction. Such a process is well beyond perturbation regime, i.e., Born limit and needs resummation which is equivalent to solving the Schrodinger equation in a reduced system. The induced Yukawa potential between the two particles alters the wavefunction of the reduced system at $r = 0$, thus resulting SE~\cite{Hisano:2006nn,Cirelli:2007xd,Feng:2010zp,Cassel:2009wt,Dent:2009bv} for relic density and indirect detection. Meanwhile it also affects the wavefunction  at $r\rightarrow\infty$,  thus enhancing the self-interacting cross section. The underlying reason for the existence of enhancements on both $r\rightarrow 0$ and $r\rightarrow\infty$ comes from the similarity of ladder diagrams between these two processes. The only difference comes from the fact that annihilation diagram needs additional operator insertion for annihilating into SM particles, so the extent of these enhancements are not the same though we can solve them simultaneously.

\begin{figure} [tb]
\begin{center}
\includegraphics[width=0.4\textwidth]{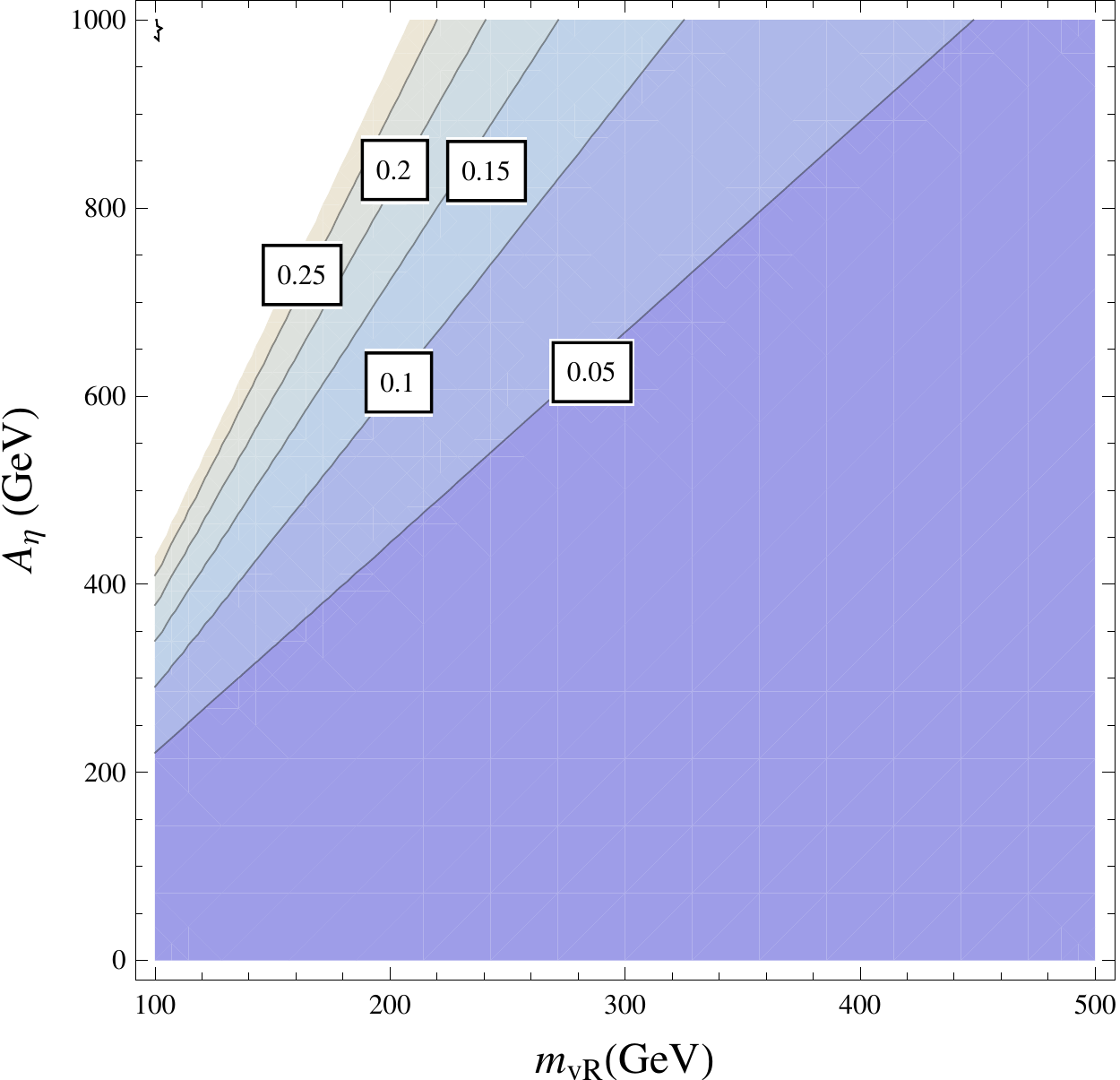}
\end{center}
\caption{The magnitude of $\alpha$ as a function of $m_{\tilde\nu_R}$ and $A_{\eta}$. }
\label{fig:alpha}
\end{figure}

Here we define a dimensionless coupling constant which plays a crucial role in non-perturbative regime as
\begin{align}
\alpha_{\eta}=\frac{1}{4\pi}\left(\frac{\lambda}{2m_{\tilde\nu_1}}\right)^2=\frac{Y_{\eta}^2}{4\pi}
\left(\frac{\sqrt{2}A_{\eta}-4v_{\eta}Y_{\eta}}{2m_{\tilde\nu_1}}\right)^2\;,
\end{align}
where $\tilde\nu_1$ is the mass eigenstate of CP-odd sneutrino which is related to soft mass $m_{\tilde\nu_R}$ and $A_{\eta}$ through equation~(\ref{eqn:mass}). In figure~\ref{fig:alpha} we give a schematic overview of the  dimensionless coupling constant $\alpha_{\eta}$ depending on $m_{\tilde\nu_R}$ and $A_{\eta}$ after setting $m_{Z^{\prime}}=1.5~\text{GeV},\, g_{BL}=0.01,\, \tan\beta^{\prime}=10$ and $Y_{\eta}=0.5$. It is easy to find at the vast parameter space, the magnitude of $\alpha$ is approximately $0.01-0.25$.
The wavefunction with spherical symmetry can be expanded into spherical harmonics,
\begin{align}
\psi(r)=\sum_{lm}R_{l}(r)Y_{lm}(\theta,\phi)\;.
\end{align}
The radial part of the wavefunctin encodes all the relevant information that needs to be solved,
\begin{align}
\frac{1}{r^2}\frac{d}{dr}\left(r^2\frac{dR_l}{dr}\right)+\left(k^2-\frac{l(l+1)}{r^2}-2\mu V(r)\right)R_l=0\;,
\label{eqn:schrodinger}
\end{align}
where $V(r)$ is long-range potential that must be given at first. The sommerfeld effect in MSSM is given in~\cite{Beneke:2012tg,Hellmann:2013jxa,Beneke:2014gja} explicitly through non-relativistic effective field theory (NREFT). The derivation is quite similar with the quarkonium except the difference that the sneutrino and bilepton are both scalars such that we do not need to worry about spin statistics. From the lagrangian equation~(\ref{eqn:lag}), the light mediator $\eta_1$ could be integrated out with effective lagrangian being given as~\cite{vandenAarssen:2013vea},
\begin{align}
S_{\text{eff}}&=\int d^4 x\left(\frac{1}{2}\partial^{\mu}\tilde\nu_1\partial_{\mu}\tilde\nu_1+\frac{m_{\tilde\nu_1}^2}{2}\tilde\nu_1^2\right)\nonumber\\
                  & +\frac{i}{2}\int d^4 x d^4 \,y j(x) D_{\eta_1}(x-y)j(y)\;,
\end{align}
where the current $j(x)$ is defined to be $\lambda\tilde\nu_1^2/2$ and $D_{\eta_1}$ is the propogator of $\eta_1$. The residue field $\tilde\nu_1$ is further split up into non-relativistic and relativistic part in momentum space. The induced potential $V$ is equivalent to the interaction term with only non-relativistic fields,
\begin{align}
V&=\frac{i}{2}\int d^4 x d^4 y j(x) D_{\eta_1}(x-y)j(y)\nonumber\\
  &=i\frac{\lambda^2}{8}\int d^4 x d^4 y\, \tilde\nu_{1\text{NR}}^{2} D_{\eta_1}(x-y)\tilde\nu_{1\text{NR}}^2\;.
\label{eqn:potential}
\end{align}
The leading order contribution of sneutrino annihilation into bileptons can be obtained from the imaginary part of box diagrams for self-scattering i.e. optical theorem. Therefore, the corresponding effective action is
\begin{align}
S_{\text{ann}}=&-i\frac{A^4}{8}\int d^4 x d^4 y d^4 z d^4 w (\tilde\nu_{1\text{NR}}  D_{\eta_1}(x-y)
\tilde\nu_{1\text{NR}}\nonumber\\&D_{\tilde\nu_1}(y-z)\tilde\nu_{1\text{NR}}D_{\eta_1}(z-w)\tilde\nu_{1\text{NR}}
D_{\tilde\nu_1}(w-z))\nonumber\\
&+(z\rightarrow w)\;.
\label{eqn:annihilation}
\end{align}
Here the two different terms are known to be t-channel and u-channel box diagrams respectively. The non-relativistic form of sneutrino is given as
\begin{align}
\tilde\nu_{1\text{NR}}=\frac{1}{\sqrt{2m_{\tilde\nu_1}}}\left(\phi(x)\exp(-im_{\tilde\nu_1}t)+\phi(x)^{\dagger}
\exp(im_{\tilde\nu_1}t)\right)\;.
\label{eqn:nonrela}
\end{align}

Combining equations~(\ref{eqn:potential},\ref{eqn:annihilation}) with (\ref{eqn:nonrela}) yields the potential and annihilation cross section,
\begin{align}
V&=-\frac{\lambda^2}{16\pi m_{\tilde\nu_1}^2}\frac{e^{-m_{\eta_1}r}}{r}=-\frac{\alpha_{\eta}}{r}e^{-m_{\eta_1}r}\;,\nonumber\\
(\sigma v_{\text{rel}})_0&=\frac{3\lambda^4}{512\pi m_{\tilde\nu_1}^6}\frac{(1-m_{\eta_1}^2/m_{\tilde\nu_1}^2)^{1/2}}{[1-m_{\eta_1}^2/4m_{\tilde\nu_1}^2]^2}\nonumber\\
&=\frac{3\pi\alpha_{\eta}^2}{2m_{\tilde\nu_1}^2}\frac{(1-m_{\eta_1}^2/m_{\tilde\nu_1}^2)^{1/2}}{[1-m_{\eta_1}^2/4m_{\tilde\nu_1}^2]^2}\;,
\label{eqn:result}
\end{align}
where in first line of equation~(\ref{eqn:result}), we recognize the Yukawa potential. The negative sign means the effective potential of scalar-scalar annihilation mediated by a light scalar particle is always attractive. The second line represents the annihilation cross section and the subscript $0$ denotes that it does not consider loop and non-perturbative correction.  It is also responsible for the finite lifetime of the bound state~\cite{Petraki:2015hla,Petraki:2016cnz} due to the decay of the reduced system where bound state itself has an impact on dark matter freeze-out. Since our focus in this paper is to demonstrate the long-range effect of the bilepton $\eta_1$, we do not consider the bound state effects here and leave it in the future work. A different, but completely equivalent approach~\cite{Iengo:2009ni,Buckley:2009in,Hryczuk:2010zi,Hryczuk:2011tq,Liu:2013vha} can reproduce the Yukawa  potential. But the advantage of NREFT is that it not only calculate potential but obtain annihilation cross section. That is why we adapt this approach in this paper.

Schrodinger equation with Yukawa  potential can not be solved in a analytical form. The partial wave expansion method is thus introduced to cope with it numerically. The starting point is the assumption of wave function,
\begin{align}
\psi(r)&\sim e^{ikz}+f(\theta) \frac{e^{ik\cdot r}}{r}\;,\nonumber\\
f(\theta)&=\sum_{l=0}^{\infty}(2l+1)f_l(k)P_{l}(\cos\theta)\;,
\end{align}
where $P_{l}(\cos\theta)$ is the Bessel function, and $f_l(k)$ encodes the scattering information,
\begin{align}
f_l(k)=\frac{e^{2i\delta_l(k)}-1}{2ik}\;.
\end{align}
The one that determines the scattering process is the phase shift $\delta_l$ of a certain partial wave. From $\delta_l$ we can obtain the differential cross section,
\begin{align}
\frac{d\sigma}{d\Omega}=\frac{1}{k^2}\left|(2l+1)e^{i\delta_l}P_l(\cos\theta)\sin\delta_l\right|^2\;,
\end{align}
The integral viscosity cross section is expressed in terms of phase shifts $\delta_l$, via the analytical non-relativistic formula,
\begin{align}
\sigma_V=\frac{4\pi}{k^2}\sum_{l=0}^{\infty}\frac{(l+1)(l+2)}{(2l+3)}\sin^2 (\delta_{l+2}-\delta_{l})\;,
\end{align}

In order to obtain $\sigma_V$, one must solve for $\delta_l$ from the asymptotic behavior of radial function,
\begin{align}
\lim_{r\rightarrow\infty} R_l(r)\sim \cos\delta_l j_l(kr)-\sin\delta_l n_l(kr)\;,
\end{align}
where $j_l$ and $n_l$ are the spherical Bessel and Neumann functions, respectively. We explicitly calculate the viscosity cross section $\sigma_{V}$ from $\delta_l$ numerically. The equation~(\ref{eqn:schrodinger}) is not good to handle in numerical implemention. We adapt the following parameter choice~\cite{Buckley:2009in,Tulin:2013teo}:
\begin{align}
\chi_l&=r R_l\;,\quad x=\alpha_{\eta}m_{\tilde\nu_1}r\;,\nonumber\\
a&=\frac{v}{2\alpha_{\eta}}\;,\quad b=\frac{\alpha_{\eta}m_{\tilde\nu_1}}{m_{\eta_1}}\;.
\label{eqn:parameter}
\end{align}
In terms of the equation~(\ref{eqn:parameter}), the radial wavefunction is reduced to be
\begin{align}
\left(\frac{d^2}{dx^2}+a^2-\frac{l(l+1)}{x^2}+\frac{1}{x}\exp(-x/b)\right)\chi_l=0\;.
\end{align}
Before showing our numerical results, we briefly discuss the effects of various  constraints for SIDM on our model:
\begin{itemize}
\item Thermal relic abundance should yields $\Omega_{\tilde\nu_1}h^2\simeq0.12$ and the impact of the SE on the leading order annihilation cross section in equation~(\ref{eqn:result}) should be taken into account. However, naturalness argument requires the sneutrino mass not to be larger than $100~\text{GeV}$ unless bileptino becomes LSP. Thus SE at freeze-out is negligible. We can use relic density requirement to determine $\alpha_{\eta}$ with $m_{\tilde\nu_1}$, $m_{\eta_1}$ being input parameters.
\item Bilepton mediator decay provides a robust constraint on sneutrino annihilation process from CMB observation and indirect detections. The enhancement at recombination time exclude most the parameter space of DM annihilation dominated by s-wave process. In BLSSM, the story is slightly different where bilepton mediator is not directly decaying into SM final states but into right-handed neutrinos. Since the right-handend neutrino must be lighter than the mediator $\eta$ in order to allow (\ref{eqn:BBN}) kinematically accessible, its subsequent decay process must be off-shell into three fermions~\cite{GonzalezGarcia:1990fb,Escudero:2016tzx}. Such decay width is so small due to the three body phase space suppression that right-handed neutrino could be served as a long-live intermediator~\cite{Kim:2017qaw,Chu:2017vao}. As the effective DM density that enters in the indirect signals is smeared by mediators, observations from indirect detection especially gamma rays are much less constraining than that of \cite{Bringmann:2016din} where the vector mediator SIDM suffers from indirect detection constraints seriously. Furthermore, the right-handed neutrino at recombination scale does not insert energy for CMB, thus the retarded decay process also solve the CMB issue. One thus expects there is no CMB and indirect detection constraints over SIDM parameter space. For the detailed discuss on this issue, we will leave for another work.
\end{itemize}

As a consequence, the only two important constraints in our model are relic abundance and viscosity cross section by SIDM. In our model, The dominant DM annihilation channel is $\tilde{\nu}_1\tilde{\nu}_1\to \eta_1\eta_1$, corresponding annihilation cross section is given in equation~(\ref{eqn:result}). One can further eliminate $\alpha_\eta$ through relic abundance constraint and leaving DM mass $m_{\tilde{\nu}_1}$ and mediator mass $m_{\eta_1}$ as two free parameters.

\begin{figure} [!htbp]
\begin{center}
\includegraphics[width=0.5\textwidth]{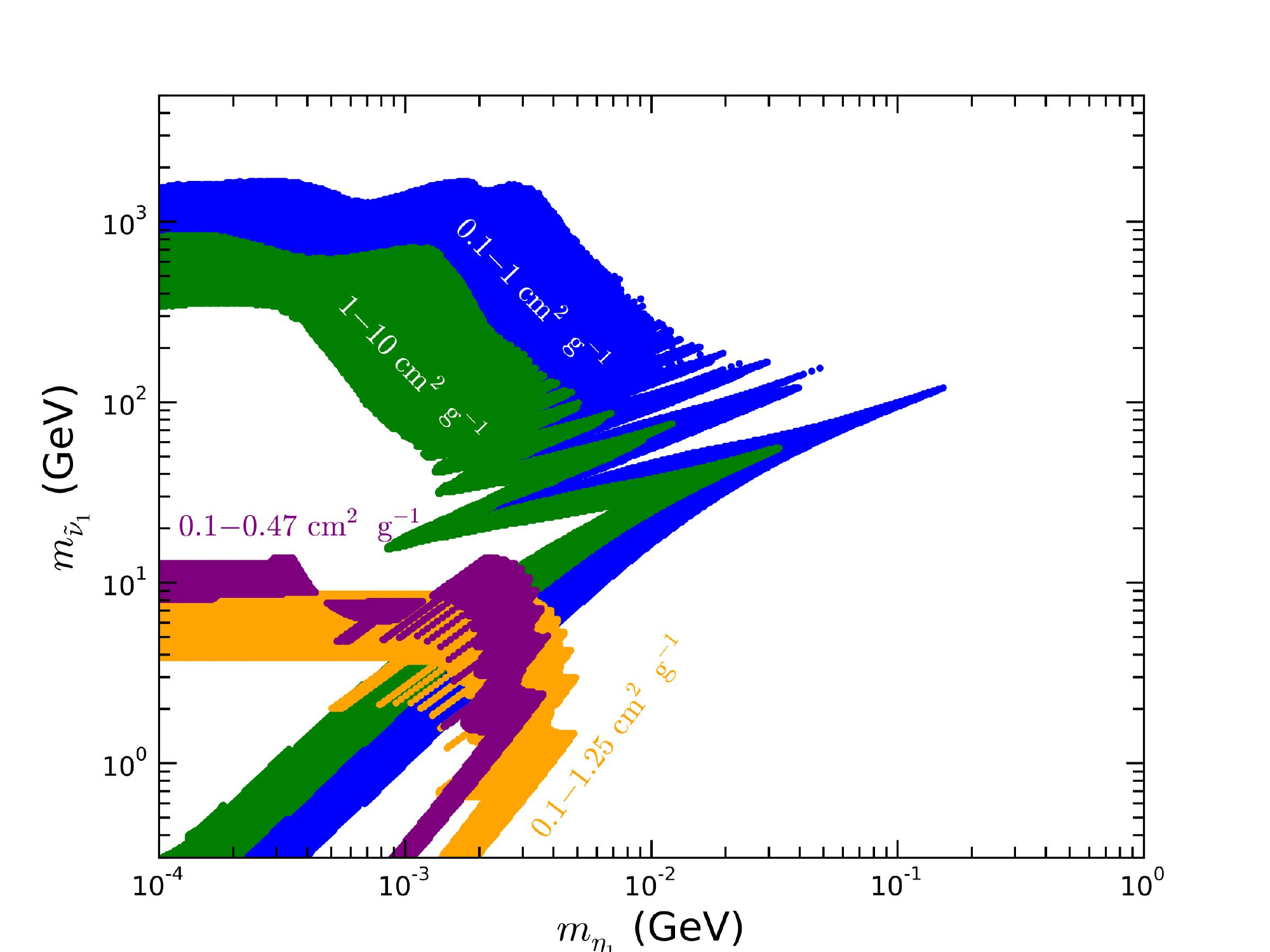}
\end{center}
\caption{Preferred parameter regions by relic abundance and SIDM in $[m_{\eta_1},~m_{\tilde{\nu}_1}]$ plane. The blue and green shaded regions corresponding to $\langle\sigma_V/m_{\tilde{\nu}_1}\rangle_{30} \in 0.1-10~{\rm cm}^2~{\rm g}^{-1}$, which would significantly affect astrophysical observables at scale of dwarf galaxies. There are also shown regions preferred by galaxy clusters with $\langle\sigma_V/m_{\tilde{\nu}_1}\rangle_{2000} \in 0.1-1.25~{\rm cm}^2~{\rm g}^{-1}$ (orange shaded) and $\langle\sigma_V/m_{\tilde{\nu}_1}\rangle_{900} \in 0.1-0.47~{\rm cm}^2~{\rm g}^{-1}$(purple shaded).}
\label{fig:dwarf}
\end{figure}

Here we follow Ref.~\cite{Kaplinghat:2015aga} to investigate effect of SIDM on different galaxy scale. For this purpose, we define $\langle\sigma_{V}/m_{\tilde\nu_1}\rangle_v$ as $\sigma_{V}$ averaged over a Maxwellian velocity distribution with the most probable velocity equals to $v~({\rm km}~{\rm s}^{-1})$ which corresponding to typical velocity dispersion for given galaxy. For dwarf galaxies, we choose $\langle\sigma_V/m_{\tilde{\nu}_1}\rangle_v \in 0.1-10~{\rm cm}^2~{\rm g}^{-1}$ at $v=30~{\rm km}{\rm s^{-1}}$. While for galaxy clusters, there exist several upper limits: the constraint from bullet cluster requires $\sigma_V/m_{\tilde{\nu}_1} < 1.25 {\rm cm}^2~{\rm g}^{-1}$ at $68\%$ confidential level (C.L.)~\cite{Randall:2007ph} and the constraint from ensemble of merging clusters imposes $\sigma_V/m_{\tilde{\nu}_1} < 0.47 {\rm cm}^2~{\rm g}^{-1}$ at $95\%$ C.L.~\cite{Harvey:2015hha}. We thus choose more conservative regions that $\langle\sigma_V/m_{\tilde{\nu}_1}\rangle_v \in 0.1-1.25~{\rm cm}^2~{\rm g}^{-1}$ at $v=2000~{\rm km}~{\rm s^{-1}}$ and $\langle\sigma_V/m_{\tilde{\nu}_1}\rangle_v \in 0.1-0.47~{\rm cm}^2~{\rm g}^{-1}$ at $v=900~{\rm km}~{\rm s^{-1}}$, respectively. Finally, for relic abundance, we require it matches the Planck
observed value $\Omega_{\tilde\nu_1}h^2=0.1199\pm0.0027$ at $95\%$ C.L.~\cite{Ade:2015xua}. Our results shown in figure~\ref{fig:dwarf} illustrate the preferred parameter regions for different galaxy scales. The blue and green shaded
bands show the regions preferred by dwarf galaxies, while the orange and purple regions present the parameter space preferred by galaxy clusters. Comparing with dwarf galaxies, galaxy clusters hold smaller regions due to more stringent requirement on $\sigma_V/m_{\tilde{\nu}_1}$, and the common parameter space can exist for $m_{\tilde{\nu}_1}\sim 1-10~{\rm GeV}$ and $m_{\eta_1}\sim 1-4~{\rm MeV}$. Moreover, for coupling $\alpha_\eta$, one observe from equation~(\ref{eqn:result}) that annihilation cross section is insensitive to mediator mass for $m_{\eta_1}\ll m_{\tilde{\nu}_1}$. $\alpha_\eta$ is thus increase monotonously with DM mass $m_{\tilde{\nu}_1}$ and varying from $5\times10^{-6}$ to $3\times10^{-2}$ (from $5\times10^{-6}$ to $3\times10^{-4}$) for dwarf galaxies (galaxy clusters) within our interested parameter regions.

%\begin{figure} [!htbp]
%\begin{center}
%\includegraphics[width=0.5\textwidth]{bullet_small.pdf}
%\end{center}
%\caption{Similar with figure~\ref{fig:dwarf}, but for bullet cluster scale. The cyan and pink shaded regions corresponding to $\langle\sigma_V/m_{\tilde{\nu}_1}\rangle_{2000} \in 0.1-1.25~{\rm cm}^2~{\rm g}^{-1}$ and $\langle\sigma_V/m_{\tilde{\nu}_1}\rangle_{900} \in 0.1-0.47~{\rm cm}^2~{\rm g}^{-1}$, respectively.}
%\label{fig:bullet}
%\end{figure}

\section{Conclusion}
\label{sec:conclusion}

In this paper, we have discussed in detail the realization of SIDM within the framework of $U(1)_{B-L}$ extension of MSSM and DFP mechanism. The right-handed sneutrino serve as LSP in most of parameter space for non-vanishing soft trilinear coupling $T_{\eta}$. It is thus compatible with the requirement of DFP SUSY without introducing dangerous direct detection limits. Its relic abundance is achieved via sneutrino annihilate into pair of bilepton which also serve as light mediator in DM self-scattering. More interestingly, annoying CMB constraint can be escaped by  retarded decay of right-handed neutrinos. The numerical calculation indicates that SIDM can be realized in our model from the scales of dwarf galaxy to galaxy cluster.

\begin{appendix}
\section{Derivation of mass splitting in bileptino}
\label{app:splitting}

Here we give a derivation of the mass splitting for bileptinos in BLSSM. The procedure is similar to~\cite{Nagata:2014wma,Krall:2017xij}. In BLSSM, the mass term for bileptinos $\tilde\eta_{1,2}$ is given explicitly from superpotential,
\begin{align}
\mathcal{L}_{bileptino}=-\mu_{\eta}\tilde\eta_1\tilde\eta_2+{\rm h.c.}\,.
\end{align}
$\tilde\eta_{1,2}$ is Dirac fermion if no mixing effect is introduced. As a result an accidental global $U(1)$ symmetry is induced where $\tilde\eta_{1,2}$ could be rotated with each other. The mixing effect which divides Dirac fermion into two nearly degenerated Majorana fermion thus must come from $U(1)$ breaking operator. Considering the dimension-five operator,
\begin{align}
\mathcal{L}_{\text{eff}}=\sum_{i=1}^2 c_i \mathcal{O}_i+h.c.\;,
\end{align}
where
\begin{align}
\mathcal{O}_1&=\eta_2 \tilde\eta_1 \eta_2\tilde\eta_1\;,\nonumber\\
\mathcal{O}_2&=\eta_1 \tilde\eta_2 \eta_1\tilde\eta_2\;.
\end{align}
These two operators can be obtained through integrating out the corresponding heavy BLino which is the superpartner of the $U(1)_{B-L}$ gauge boson. The wilson coefficients give rise to mass splitting for bileptino that can be found in the mass matrix directly,
\begin{align}
\mathcal{L}=-\frac{1}{2}(\tilde\eta_1,\tilde\eta_2)\mathcal{M}
\left(
\begin{array}{c}
 \tilde\eta _1 \\
 \tilde\eta _2
\end{array}
\right)\;,
\end{align}
with
\begin{align}
\mathcal{M}=\left(
\begin{array}{cc}
 -\cos^2\beta^{\prime} c_1 v_{\eta }^2 & -\mu
   _{\eta } \\
 -\mu _{\eta } & -\sin^2\beta^{\prime} c_2
   v_{\eta }^2
\end{array}
\right)\;.
\end{align}
The symmetric mass matrix $\mathcal{M}$ can be diagonalized and the result mass eigenvalues are
\begin{align}
m_1&=\mu-\frac{1}{2}(c_1 \cos^2\beta^{\prime}+c_2\sin^2\beta^{\prime}) v_{\eta}^2\;,\nonumber\\
m_2&=\mu+\frac{1}{2}(c_1 \cos^2\beta^{\prime}+c_2\sin^2\beta^{\prime}) v_{\eta}^2\;.
\end{align}
Therefore, the mass splitting is found to be
\begin{align}
\delta=m_2-m_1=(c_1 \cos\beta^{\prime}+c_2\sin\beta^{\prime}) v_{\eta}^2\;.
\end{align}
The Wilson coefficients $c_1$ and $c_2$ can be obtained by matching the blino-bleption-bilepton vertex,
\begin{align}
c_1=c_2=\frac{g_{\text{BL}}^2}{4M_{B^{\prime}}^2}\;.
\end{align}
The resulted mass splitting is
\begin{align}
\delta=\frac{m_{Z^{\prime}}^2}{2M_{B^{\prime}}}\;.
\end{align}
\end{appendix}

\section*{Acknowledgements}
We would like to thank Wenyu Wang for help on numerical calculation, and Fei Wang for helpful discussion. Bin Zhu is supported by the National Science Foundation of China (11747026 and 11575151). YLi is also supported by the Natural Science Foundation of Shandong Province (Grant No.ZR2016JL001).

\bibliography{lit}

\end{document}